\documentclass[conference]{IEEEtran}
\IEEEoverridecommandlockouts
\usepackage{cite}
\usepackage{color,soul}
\usepackage{amsmath,amssymb,amsfonts}
\usepackage{graphicx}
\usepackage{textcomp}
\usepackage{xcolor}
\usepackage{braket}
\usepackage{algorithm}
\usepackage{algpseudocode}

\usepackage[a4paper, total={184mm,239mm}]{geometry}
\def\BibTeX{{\rm B\kern-.05em{\sc i\kern-.025em b}\kern-.08em
    T\kern-.1667em\lower.7ex\hbox{E}\kern-.125emX}}
\begin{document}

\title{Forensics of Transpiled Quantum Circuits\\
}

\author{
    \IEEEauthorblockN{Rupshali Roy}
    \IEEEauthorblockA{
        \textit{School of EECS} \\
        \textit{Penn State University} \\
        University Park, United States \\
        rzr5509@psu.edu
    }
    \and
    \IEEEauthorblockN{Archisman Ghosh}
    \IEEEauthorblockA{
        \textit{School of EECS} \\
        \textit{Penn State University} \\
        University Park, United States \\
        apg6127@psu.edu
    }
    \and
    \IEEEauthorblockN{Swaroop Ghosh}
    \IEEEauthorblockA{
        \textit{School of EECS} \\
        \textit{Penn State University} \\
        University Park, United States \\
        szg212@psu.edu
    }
}

\maketitle

\begin{abstract}

 There is a steady growth in research and demand for real quantum hardware in the Noisy Intermediate-Scale Quantum (NISQ) era of quantum computing. This prompts many third-party cloud providers to set up quantum hardware as a service that includes a wide range of qubit technologies and architectures to maximize performance at minimal cost. Different backends vary in terms of noise behavior, the basis gate set, coupling architecture and speed, among other factors. The user has the flexibility to choose from various quantum hardware, qubit technologies, and coupling maps; however, there is little to no visibility on where the execution of the circuit is actually taking place. This situation is similar to classical cloud which may include processors from various vendors and the user programs can be stored and executed on any of them. The difference in the quantum scenarios is that the success of the user program (among various other metadata of the program execution like performance, number of iterations to convergence for variational algorithms and cost) is highly reliant on the backend that was used for execution. Besides, the third-party provider and/or tools (e.g., hardware mapper and allocator) may be untrustworthy and execute the quantum circuits on less efficient and more error-prone hardware to conserve resources and maximize profit. As such, gaining visibility on the backend from various aspects of the computing e.g., transpilation, execution and outcomes will be extremely valuable. Towards this goal, we introduce the problem of forensics in the domain of quantum computing where the objective is to trace the hardware and its properties where the quantum program has been transpiled and executed. Effective forensics can have many applications including establishing trust in the quantum cloud services. In this work, we focus on tracing the coupling map of the hardware (from the transpiled program) where the transpilation of the circuit took place. We perform experiments on various coupling topologies (linear, T-shaped, H-shaped, and loop) on IBM backends. We are able to derive the coupling map from the transpiled circuits with complete accuracy for almost every transpiled circuit. Next, we extracted the coupling map of the whole backend using as little as 3 transpiled circuits, both for user-based and transpiler-based (using auto-assignment feature) logical to physical mapping of the qubits. Finally, we addressed the problem of tracing the backends in a suite of backends using a pool of transpiled quantum circuits of varied sizes. We could correctly trace 97.33\% of the programs to the correct backend.  

\end{abstract}


\section{Introduction}
The research in quantum computing is continually growing, and a major catalyst to this growth is cloud-based quantum computing services. As we progress through the Noisy Intermediate-Scale Quantum (NISQ) era, access to quantum hardware resources will become more abundant enabling better research and development at a lower investment. The cloud providers incorporate multi-qubit processors with different layouts, architectures, and qubit technologies, giving the circuit designers a wide variety of choice. 
However, a certain degree of opacity exists between the user and the hardware. For example, multiple isomorphic choices of coupling maps (in terms of noise, architecture, and execution speed, to name a few) exist within a large backend or among a suite of backends where the program could be mapped even though the backend, the coupling map, and qubits are specified by the user. This could be possible due to mapping and hardware allocation policy which is geared towards objectives such as maximizing hardware throughput and profit, to name a few. However, there is no means for the user to ascertain that the circuit execution occurred on the coupling map and the quantum hardware that was specified during compilation. While the cloud vendor may not consider this a serious violation of trust, the success of the program and the execution speed still relies heavily on the backend and even a slight change can lead to significant impact on optimization quality. For example, the hardware allocator might aggregate the gate error, readout error and crosstalk of various coupling map choices of the backend to find a fair and available map \cite{b0} however, the program could be asymmetrically sensitive to individual errors and may be penalized heavily in terms of computation quality and/or latency. This aspect becomes more challenging under multi-tenant environment where multiple jobs share the same hardware. A community-based partitioning of hardware is proposed \cite{b01} to allocate the physical qubits to the user programs from the queue. Appropriate metric-based measures are taken to identify malicious jobs that could inject fault into the victim's program through crosstalk and excluded from parallel execution. However, such security-aware approaches are not usually considered in conventional schedulers and allocators. Furthermore, metrics-based methods \cite{b01} could still fail to fully protect programs from intentional or unintentional noise injection from other programs sharing the hardware. Since the compilation, allocation and scheduler policies and metrics used within the tools are proprietary and protected, there is no way to validate if the decision made by such tools will meet the objectives of the users. This is in contrast to classical cloud computing where rerouting of programs to various physical hardware and virtualization (again opaque to the user) does not affect the computation quality. 

While the software stack including the transpiler (including mapping and routing process), hardware allocator, and job scheduler is developed in-house, there is a trend towards incorporating third-party tools to cut down development time and resources \cite{b12}. Under both scenarios (i.e., in-house and 3rd party developed tools), an update of the toolchain or a bug can affect the user's jobs. The involvement of untrusted parties or individuals in the tool development can exacerbate the problem of misrouting/misallocation of user program further. The adversary motivation to execute the circuit on less efficient/error-prone qubits while charging the user for high-quality qubits would include conserving quantum resources, increasing profit margin, degrading the computation quality (i.e., cause denial-of-service) and injecting failures, to name a few. 
The possibility of the tool's metric-guide algorithm's "unintentional"  poor hardware allocation and adversarially motivated "intentional" hardware misallocation of jobs is further amplified with untrusted quantum cloud vendors (i.e., quantum cloud located in and/or quantum hardware procured from less trustworthy or untrustworthy locations) where the integrity of the software toolchain is also questionable. 

The above challenges require some form of forensics (with similarity to reverse engineering) to gain insights about various decisions made within the cloud end from various observables from the job. Some examples of the observables could be transpiled programs, job execution speed and convergence time (for iterative algorithms), to name a few. At the very least, such forensics could shed light on validating the claims made by the vendor, provide insights on the metrics and policies employed by the tools and establish trust in the process. A forensics of quantum system could open up many avenues namely, (a) \emph{Fingerprinting of hardware and details:} It will be feasible to identify the hardware and its characteristics such as error rates and gate execution times where the job was executed. (b) \emph{Establishing trust in quantum system:} It will enable the user to validate whether their desired objectives were met and build trust in the process. (c) \emph{Finding bugs and/or loopholes in the tools:} It will not only facilitate the detection of bugs in the toolchain but also find loopholes such as the allocation of two jobs in parallel that inject crosstalk-induced faults into each other. 

\subsection{Forensics in digital vs quantum systems}
Digital forensics plays a crucial role in the investigation and analysis of digital evidence to ensure security and trust in the computing system \cite{b13}. Taking a page out of hardware security and trust, we find that forensic frameworks such as side-channel analysis \cite{b14}, logic testing \cite{b15} and circuit reverse engineering \cite{b16} enable researchers to validate hardware systems, investigate tampering, and build a root-of-trust. However, forensics is not relevant to identifying the hardware that is used to execute a job in the cloud setting since all machines are nearly identical and may only affect the job completion time (which in turn depends on numerous other factors such as queue depth and job priority). However, the mere task of identifying the hardware used to execute a job in a quantum cloud with a plethora of backends and qubit technologies is non-trivial. Some general challenges are, (a) the presence of noise that affects the qubit operations heavily complicating the implementation of traditional forensic ideas, (b) the inaccessible internal state of the qubits which makes the quantum hardware a black-box to the user, (c) existence of many isomorphic choices of mapping. 

\subsection{Forensics vs Fingerprinting}
The differentiating factors between forensics and fingerprinting are as follows: (a) the scope of objectives for forensics is much wider than hardware fingerprinting. For example, in addition to identifying the hardware used for execution of the job (which is the prime objective of fingerprinting), we would also like to know the policies and metrics used in the software toolchain as part of forensics, (b) fingerprinting can employ hand-crafted jobs whose objectives are solely (with no real functionality) to amplify the hardware-specific properties such as noise to aid in identification whereas forensics scavenges information from the observables of the job (with real functionality) execution. As such, forensics can be considered extremely challenging compared to fingerprinting, (c) fingerprinting can only offer limited visibility on the workings of quantum cloud whereas forensics can offer rich information. 
\subsection{Contributions}

In this paper, we demonstrate a first step towards forensics of quantum systems by understanding the transpilation process as a test case. Our objective is to extract the coupling map of the quantum hardware by analyzing the transpiled copies of utility programs\footnote{Typically hardware providers publish coupling map of the backends. However, the program may not be executed on the promised backend and/or a subgraph of the backend but a completely different backend and a subgraph.}. From the transpiled program, we also identify the backend (from a pool of backends) where the program has been mapped. We assume that the user will submit the job and let the cloud vendor perform the transpilation and mapping to the user selected backend and provide the computation results back along with the transpiled program. However, user will not have means to validate if the transpilation occurred on their desired target backend and coupling map. The proposed forensics will assist in meeting the above gap. 

We note that, transpilation of a program and mapping to a coupling map often leaves some signature e.g., a direct 2-qubit gate like CNOT between qubits indicate direct physical connectivity whereas presence of SWAP gates before the 2-qubit gate indicate that the qubits are disconnected. The length of routing path (i.e., the number of SWAP gates between the CNOT) also indicates the distance between the physical qubits. We also note that transpiled QASM code reveals the physical qubit numbers of the backend where the logical qubits are mapped. An example is shown in Fig. \ref{heuristic}. By leveraging these hints we aim to determine the coupling map of the hardware. Since the backend is typically larger than the program size, a single program can only reveal the partial architecture of the backend. We analyze many transpiled programs to reveal the whole coupling archotecture of the backend. This approach is valid under the scenario that the user submits many jobs (of various sizes) to the quantum cloud with target hardware. Therefore, they will have access to more than single transpiled copies which can aid in forensics.
We validate the efficacy of the proposed ideas by tracing the coupling for multi-qubit circuits transpiled on various IBM hardware. 
\emph{To the best of our knowledge, this is the first attempt to perform forensics on quantum systems.}


In the remainder of the paper, Section II provides background on quantum computing and related works. The coupling map extraction procedure, the experimental setup and results are presented in Section III. Conclusions are drawn in Section IV.

\section{Background}
\subsection{Preliminaries} 
\subsubsection{Qubits} Qubits are the fundamental units of quantum information. A qubit is typically represented using a two-level system, with the basis states denoted as $\ket{0}$ and $\ket{1}$. Unlike classical bits, which can only be in one state (0 or 1) at a time, qubits can exist in a superposition of these two states. The general state of a single qubit is described by a linear combination $\psi$ = $\alpha$ $\ket{0}$ + $\beta$ $\ket{1}$, where $\alpha$ and $\beta$ are complex coefficients. The squares of the magnitudes of $\alpha$ and $\beta$ give the probabilities of measuring the qubit in the $\ket{0}$ or $\ket{1}$ state, respectively, with the sum of these magnitudes equal to 1.

\subsubsection{Quantum gates} Quantum gates manipulate qubits to change their states, entangle qubits, or generate superpositions. They are represented by unitary matrices that perform transformations on quantum states, satisfying the condition $\mathbf{U}^\intercal U = I$, where $\mathbf{U}^\intercal$ is the conjugate transpose of $U$, and $I$ is the identity matrix. Common quantum gates include the rotation gates (Rx, Ry, Rz), the Hadamard gate, the Pauli-X gate, and the controlled-NOT (CNOT) gate. Rotation gates Rx (or Ry) rotate the qubit's state vector around the x-axis (or y-axis) of the Bloch sphere by a phase given in radians. The Hadamard gate creates an equal superposition of the $\ket{0}$ and $\ket{1}$ states, while the Pauli-X gate flips the qubit's state. The CNOT gate, a two-qubit gate, flips the target qubit if the control qubit is in the $\ket{1}$ state.

\subsubsection{Basis gates and coupling constraints} In practice, quantum computers support only a specific set of single- and multi-qubit gates, known as basis gates or native gates of the hardware. For example, IBM quantum devices use native gates like the CNOT (two-qubit), u1, u2, u3, and id (single-qubit) gates. If a quantum circuit contains non-native gates (e.g., a Toffoli gate), these gates must be decomposed into basis gates before execution. Additionally, two-qubit operations like CNOT are only allowed between connected qubits, a restriction referred to as coupling constraint. If a two-qubit gate involves unconnected qubits, SWAP gates are inserted to satisfy these constraints.

\subsubsection{Compilation} Quantum circuit compilers like Qiskit transform input circuits to comply with the hardware's coupling constraints, often inserting SWAP gates to achieve this. Compilers also optimize circuits by merging, canceling, or reordering single- and multi-qubit gates, as well as performing rotations. IBM Qiskit provides support for barriers, which prevent optimizations across specific sections of the circuit. Fig. \ref{heuristic}.1 shows how the compiler transforms a given circuit by placing SWAP gates between qubits that are not physically connected on the hardware.




\subsubsection{Hamming Distance} Hamming Distance (HD) between two graphs is a metric to quantify how different they are in terms of their edge structure. In particular, it counts how many edges differ between the two graphs. 
If two graphs \( G_1 \) and \( G_2 \) have edge sets \( E_1 \) and \( E_2 \), the Hamming distance \( d_H(G_1, G_2) \) is:
\[d_H(G_1, G_2) = |E_1 \oplus E_2|\]
Where, \( E_1 \oplus E_2 \) is the symmetric difference between the edge sets, i.e., the edges that are in either \( E_1 \) or \( E_2 \), but not in both.
\( |E_1 \oplus E_2| \) is the number of edges in the symmetric difference. In this work, we use HD to measure the similarity between the actual and extracted coupling maps.


\subsection{Related works}
Although work in the domain of forensics of quantum computing is non-existent, there have been work on fingerprinting of quantum hardware. These works use a probing circuit to capture the unique error characteristics of quantum devices \cite{b1}. When the user inspects the service, the execution results of the probing circuit act as the device-side fingerprint of the quantum hardware. 
In \cite{b2}, the authors fingerprint quantum servers by running a user's circuit with two different levels of noise, using the resultant performance gap as a fingerprint. In \cite{b3}, the frequency of qubits is used to identify quantum computers based on transmon qubits by noting that the frequencies of individual qubits are unique. This is mainly due to variations in the manufacturing process, which cause distinct physical properties for each qubit. A quantum physically unclonable function (QuPUF) is used \cite{b4} to fingerprint the requested hardware on the transpiled circuit. Two flavors of QuPUF circuits are proposed- one based on superposition and one on decoherence. As pointed before, the fingerprinting techniques rely on circuits with no real functionality. Therefore, they are not relevant for forensics purposes. 

\section{Proposed Approach and Results}
This section presents a procedure for extraction of coupling map of the hardware from the transpiled program. Next, it considers various scenarios to extract the partial coupling map of the hardware from individual programs, whole coupling map of the hardware from multiple programs and trace the programs to correct hardware in a pool of backends.
\subsection{Coupling Map Extraction}\label{AA}

\begin{algorithm}
\caption{Retrieve Coupling Map}
\label{CM}
\begin{algorithmic}[1]  
\Function{derive\_coupling\_map}{$transpiled\_circuit$}
    \State $derived\_coupling\_map \gets \emptyset$
    \State $swap\_history\_matrix \gets False for all entries$
    \For{each $instruction$ in $transpiled\_circuit.data$}
        \If{\Call{is\_swap\_gate}{$instruction$}}
            \State $swap\_history\_matrix[q1][q2] \gets True$
            \State $swap\_history\_matrix[q2][q1] \gets True$
            \State \textbf{continue} //Skip to next instruction
        \EndIf
        
        \If{$instruction$ involves 2 qubits}
            \State $q1,q2 \gets$ extract qubits from $instruction$
            \If{! $swap\_history\_matrix([q2][q1] or [q1][q2])$}
                \State $derived\_coupling\_map \gets q1,q2$
            \EndIf
        \EndIf
    \EndFor
\EndFunction
\end{algorithmic}
\end{algorithm}

\begin{figure} 
    \centering
        \vspace{-3mm}
        \includegraphics[width=0.5\textwidth]{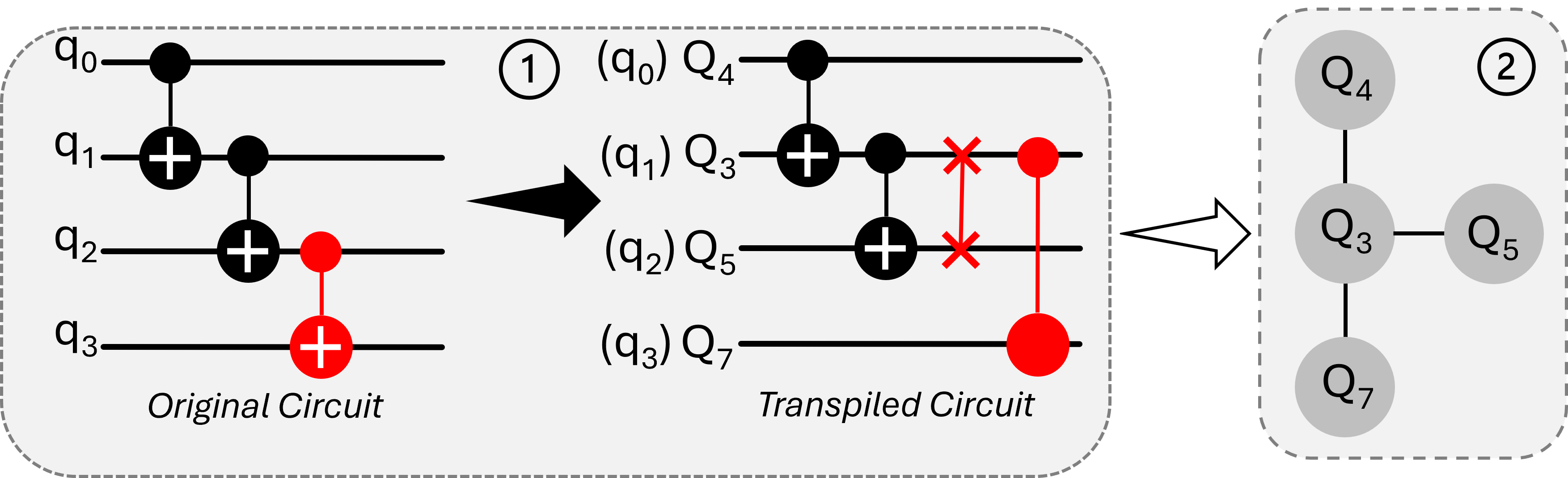}
        \vspace{0mm}
        \caption{Derivation of the physical coupling map from a transpiled circuit.}
        \label{heuristic}
\end{figure}

Based on the notion that two qubit operations are only allowed between qubits that are physically connected on the coupling map of the backend SWAP gates are inserted to execute two-qubit gate operation between unconnected qubits, we propose a heuristic (Algorithm-\ref{CM}) to extracts the coupling map from the transpiled circuit. We first iterate through the transpiled circuit to detect the presence of SWAP gates using the function $is\_swap\_gates$, which may manifest themselves in the following ways: (a) direct SWAP gate, explicitly defined as an instance of the `SwapGate' class, (b) named SWAP gate, which may be defined in a custom way, but carries out the SWAP operation and (c) gate operation for which the unitary matrix is similar to that of a SWAP gate, (d) sequence of 3 CNOT gates with qubit connections which is equivalent to a SWAP gate, (e) iSWAP gate followed by an S gate to reverse the phase from the iSWAP gate, resulting in a SWAP operation, (f) an RXX, RYY, and RZZ gate each having phase = $\pi$/2 which is equivalent to SWAP. Once a SWAP gate is detected, the state of the involved qubits is set to True in $swap\_history\_matrix$ in the corresponding entry. Note, $swap\_history\_matrix$ is a $nxn$ matrix where $n$ is the number of qubits in the program. Each entry denotes if there is a SWAP gate between the associated qubits or not. For example, the 4x4 $swap\_history\_matrix$ for the example in Fig. \ref{heuristic} will be \[
\begin{bmatrix}
0 & 0 & 0 & 0 \\
0 & 0 & 1 & 0 \\
0 & 1 & 0 & 0 \\
0 & 0 & 0 & 0 \\
\end{bmatrix}
\] 
After this tracker is updated for all the operations in the circuit, we once again iterate through them, and check which of these involve 2 qubits. If the entry corresponding to the pair of qubits involved in a 2-qubit operation have been set as True in $swap\_history\_matrix$, the qubit pair is not considered as an edge in $derived\_coupling\_map$, else it is added as an edge. After going through all of these operations, we now have a set of edges that yields the coupling map on which the circuit has been transpiled; this could either be a subgraph of the coupling map of the backend, or it may coincide with the complete coupling map of the backend if the number of qubits in the circuit and the number of qubits in the backend are equal. 

\textbf{Example: } 
In Fig. \ref{heuristic}.1, we have a 3-qubit entanglement layer with 3 $CNOT$ gates. We get the physical qubits assigned to each of the logical qubits from the transpiled circuit.qubits attribute. Since there is no direct physical connection between physical qubits $Q_5$ and $Q_7$ on the chosen backend/topology, a $SWAP$ gate is placed by the transpiler between $Q_3$ and $Q_5$ to leverage the physical connections between $Q_3$ and $Q_5$ as well as between $Q_3$ and $Q_7$. On parsing the transpiled circuit with our procedure, an initial 4 X 4 qubit swap history matrix is generated. 
The instructions are parsed in order of execution from the transpiled circuit, and the edges $Q_4-Q_3$, and $Q_3-Q_5$ (Fig. \ref{heuristic}.2) are obtained after the first two iterations since the corresponding entries are still set to 0 while we have direct operations between them. The $SWAP$ gate is encountered in the third iteration making the state of the corresponding entry in $swap\_history\_matrix$[3][5] True. In the final instruction, the operation $CNOT(Q_3, Q_7)$ is passed and the state of $swap\_history\_matrix$[3][7] being 0 in the swap history table, the edge $Q_3-Q_7$ is obtained. In this way we find all three edges giving us the coupling map for the transpiled circuit (Fig. \ref{heuristic}.2).


 
To perform a complete forensics of the backend that is being used, we transpile multiple circuits of varying sizes, and derive the coupling topology used for each from the transpiled circuit. Taking the union of these coupling topologies (in the form of sets of edges), we put together a more complete picture of the coupling map of the backend.

When the user does not define a coupling topology during transpilation and only specifies the backend for compilation and execution, the transpiler chooses qubits that are directly connected to each other for 2-qubit operations (to minimize the number of SWAP operations). The quality of qubit connections is the next factor that governs qubit selection in transpilation. 

\begin{figure} 
    \centering
        \vspace{-3mm}
        \includegraphics[width=\linewidth]{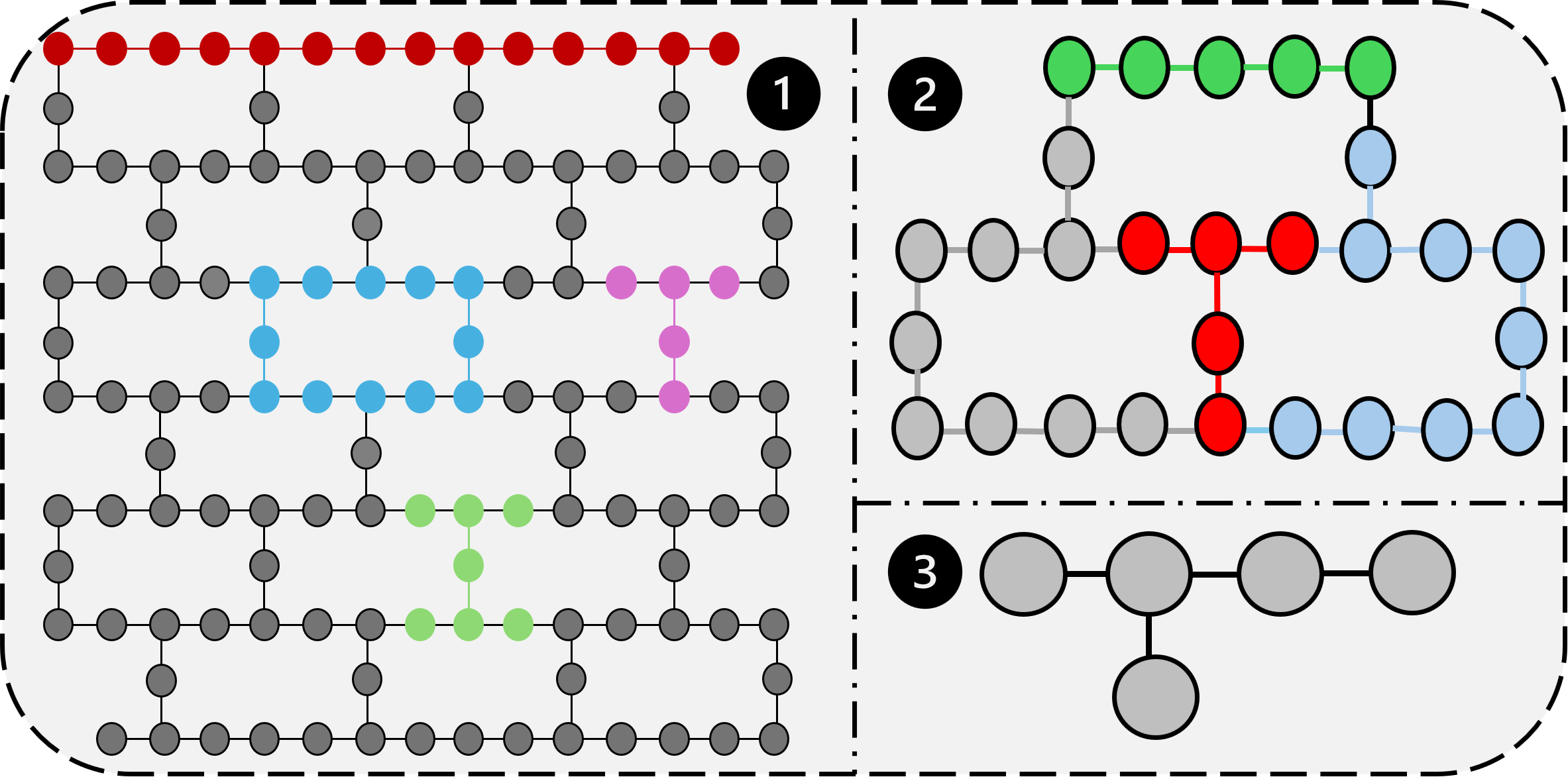}

        \caption{(1) Coupling map for IBM Kyiv. The qubits in red are an example of the linear topology, the green ones - an H-shaped topology, blue signifies the loop topology and pink is an example of the T-shaped topology. (2) Coupling map of IBM Cambridge obtained by running the analysis on the four subgraphs showed using different colors. (3) T-shaped coupling map which is identical for IBM Burlington and IBM Vigo. }
        \label{fig:cmap}
\end{figure}

\subsection{Methodology and Results}
We run the simulations using IBM Qiskit locally on an AMD Ryzen 5 5500U CPU with Radeon Graphics (2.10 GHz) machine with 8 GB RAM. We study the accuracy and time overhead (Tables \ref{tab:table} and \ref{tab:table2}) of the proposed coupling map extraction algorithm by testing it on multiple quantum circuits of varying sizes transpiled on different coupling map topologies and various scenarios. 

\subsubsection{Results for extracting coupling maps}
For a circuit with 5 qubits, we chose a 1-bit adder from Revlib \cite{b5}; for 10 qubits, a quantum Fourier transform circuit; for 15 qubits, Grover's circuit; and for 20 qubits, a GHZ state circuit. For each circuit, we considered various smaller coupling map topologies that could be supplied during transpilation on the real hardware backend, IBM Kyiv (Fig. \ref{fig:cmap}). For 5 qubits, we transpiled the circuit using a linear topology and a T-shaped map; for 10 qubits we used linear, T-shaped and H-shaped and for 15 and 20 qubit circuits we explored loop-based topologies as well. These choices are random and the extraction procedure can work equally efficiently for other chosen topologies as well. Extraction in this scenario can take from 0.21s-0.26s (Table \ref{tab:table}).


After transpilation, we ran our extraction procedure on the transpiled circuits. All the derived coupling maps matched the qubit topology that had been specified to the transpiler, barring the loop topology for 15 and 20 qubit circuits where we achieved a Hamming distance of 2 between the supplied and derived coupling maps (i.e., two edges present in the coupling map were left undetected). This shows that we are successful at correctly deriving the coupling map on which the circuit has been transpiled. 



\subsubsection{Results for extracting backend's entire coupling map} We then performed further analysis to extract the whole coupling map of the backend used to transpile circuits. A 3-qubit circuit (Miller circuit), two 5-qubit circuits (1 bit adder and a modulo 5 circuit), a 10 qubit circuit (QFT), and a 15-qubit circuit (Grover's algorithm) were transpiled on a 27 qubit fake\_cambridge backend. We then combine the derived coupling topologies together (by taking the union of the sets of edges) to extract the coupling map of whole backend.

\begin{figure} 
    \centering
        \vspace{-3mm}
        \includegraphics[width=0.9\linewidth]{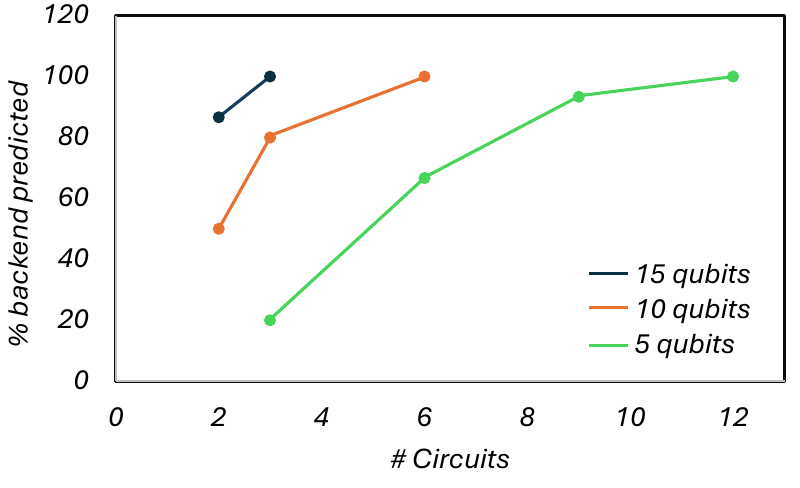}
        \vspace{0mm}
        \caption{Percentage of fake\_cambridge backend extracted with respect to the number of transpiled circuits of different sizes.}
        \label{fig:re}
\end{figure}

We also transpile multiple multi-qubit circuits of the same size (5 qubits, 10 qubits and 15 qubits) on various parts of the backend and extract the backend coupling map. We observe from Fig. \ref{fig:re} that there is a steady improvement in the percentage of coupling map of the backend predicted with an increase in the number of circuits. From Fig. \ref{fig:re} we note that the entire coupling map of the backend (100\%) can be predicted by executing as low as 12 circuits if the programs have 5 qubits each (e.g., GHZ State, Grover's circuit and QFT). For 10 qubit programs (random circuits, GHZ State, Grover's circuit and QFT circuit), 6 circuits are needed while for 15 qubit programs (GHZ State, Grover's circuit and QFT circuit) only 3 circuits are needed to extract the entire map correctly. 

\begin{figure} 
    \centering
        \vspace{-3mm}
        \includegraphics[width=0.9\linewidth]{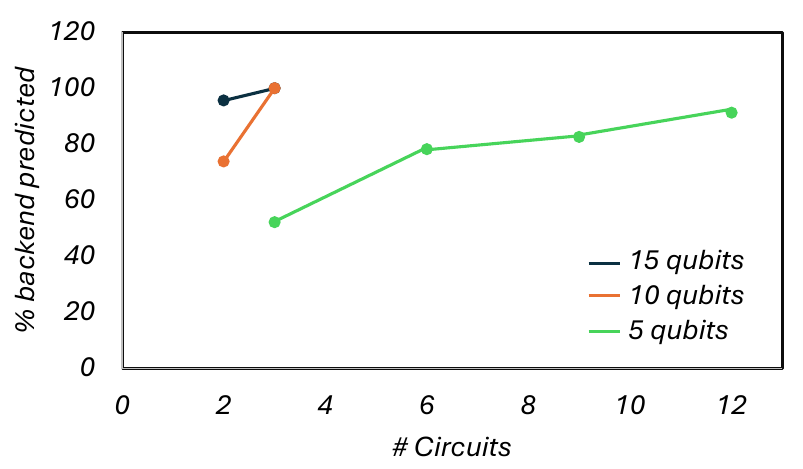}
        \vspace{0mm}
        \caption{Percentage of fake\_singapore backend extracted with respect to the number of transpiled circuits of different sizes.}
        \label{fig:re_sing}
\end{figure}
We also carried out similar analysis for 20-qubit backend named fake\_singapore (Fig. \ref{fig:re_sing}). We are able to predict 91.3\% of the backend's coupling map using 12 circuits of size 5-qubit. Using larger programs with 10 and 15 qubits, we can predict the full map with just 3 circuits.

\begin{figure}
                \centering
                    \includegraphics[width=0.46\columnwidth]{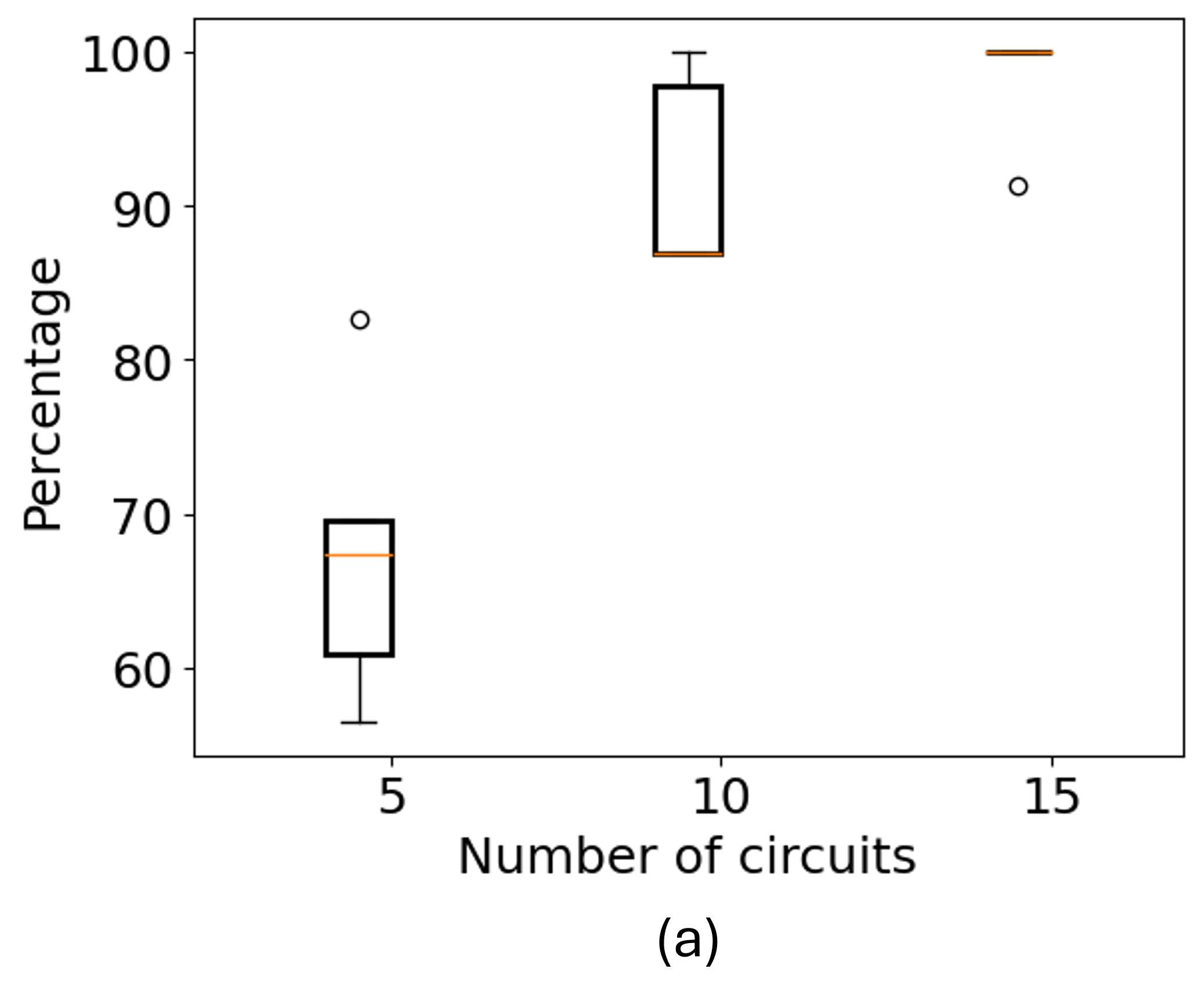}
                 \vspace{0mm}
         \label{fig:re_s_auto}
                \centering
                    \includegraphics[width=0.46\columnwidth]{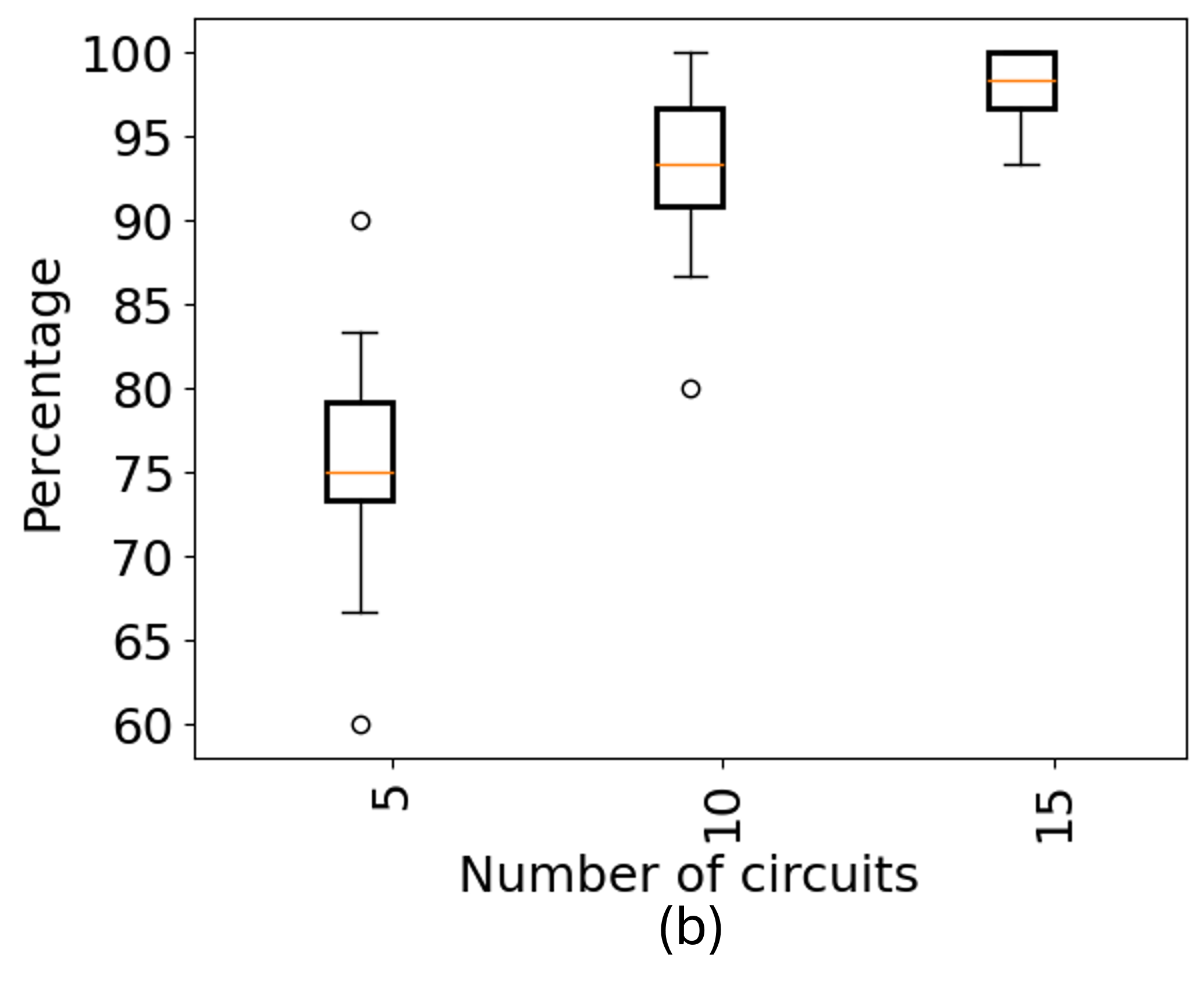}
                 \vspace{0mm}
        \label{fig:re_c_auto}
        \vspace{-1mm}
        \caption{Percentage of backend predicted using pools of differently sized transpiled circuits for (a) fake\_singapore (b) fake\_cambridge.}    
        \vspace{-3mm}
        \label{ckt_pool}
\end{figure}

Further investigation was carried out by considering 10 randomly selected pools of transpiled circuits, each of random sizes, and analyzing how much of the coupling map can be predicted with a pool of 5, 10 and 15-qubit circuits. We perform this analysis for fake\_cambridge (Fig. \ref{ckt_pool}(a)) and fake\_singapore (Fig. \ref{ckt_pool}(b)) backends. As expected, the percentage of the map predicted increases with the size of the pool. This scenario took 0.0009s for execution on both backends (Table \ref{tab:table2})

\begin{figure*}
        \centering         
        \begin{minipage}{0.46\textwidth}
                \centering
                    \includegraphics[width=\columnwidth]{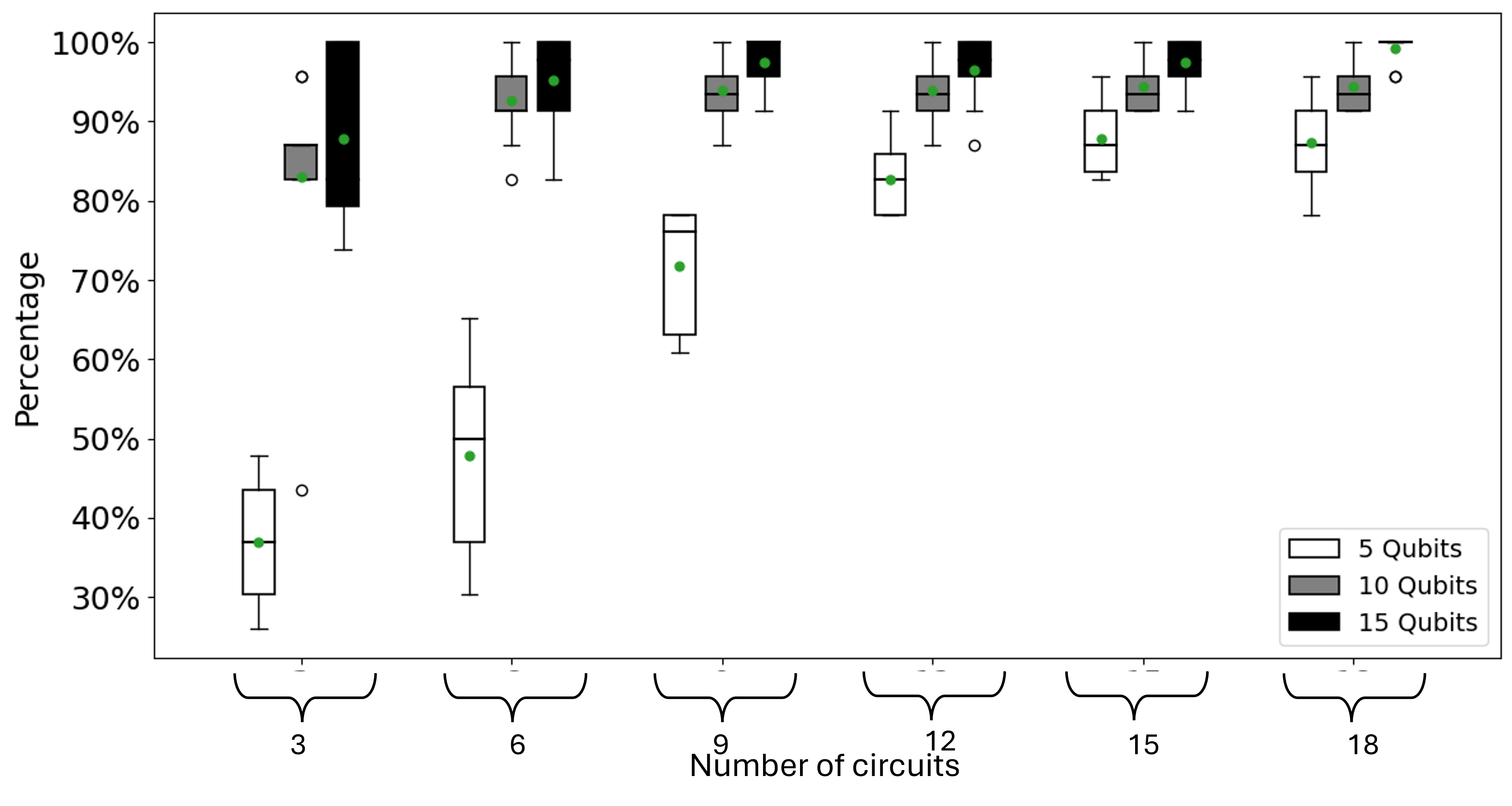}
                 \vspace{0mm}
                 (a)
         \label{fig:re_s_auto}
        \end{minipage}%
        \begin{minipage}{0.46\textwidth}
                \centering
                    \includegraphics[width=\columnwidth]{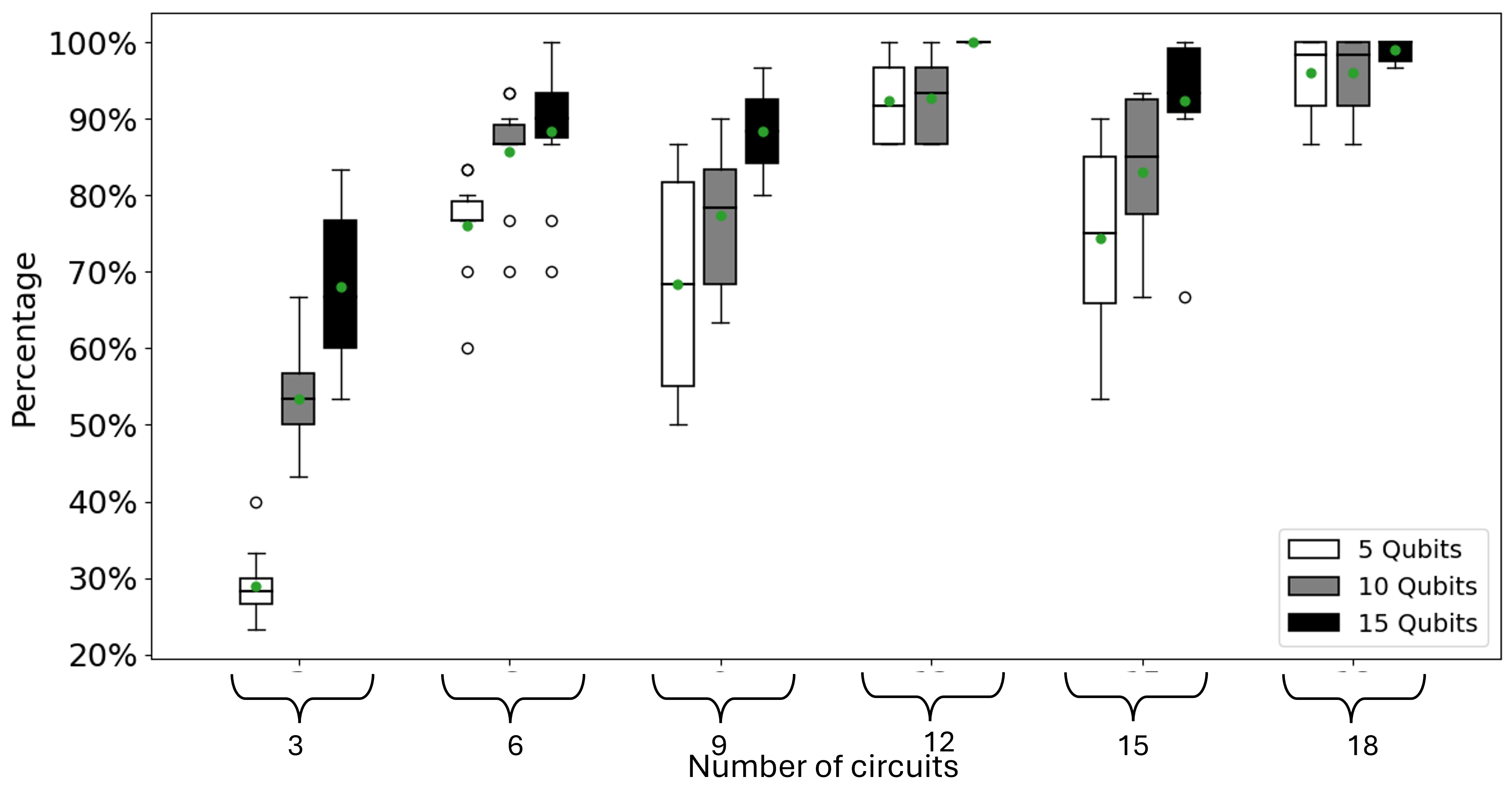}
                 \vspace{0mm}
                 (b)
        \label{fig:re_c_auto}
        \end{minipage}%
        \vspace{-1mm}
        \caption{Percentage of (a) fake\_singapore (b) fake\_cambridge backends extracted with the number of transpiled circuits of different sizes. Auto-assignment of logical-physical qubit mapping by the transpiler is used for these results. }    
        \vspace{-3mm}
        \label{re_mixed_auto}
\end{figure*}

\begin{figure}
        \centering         
                    \includegraphics[width=0.46\columnwidth]{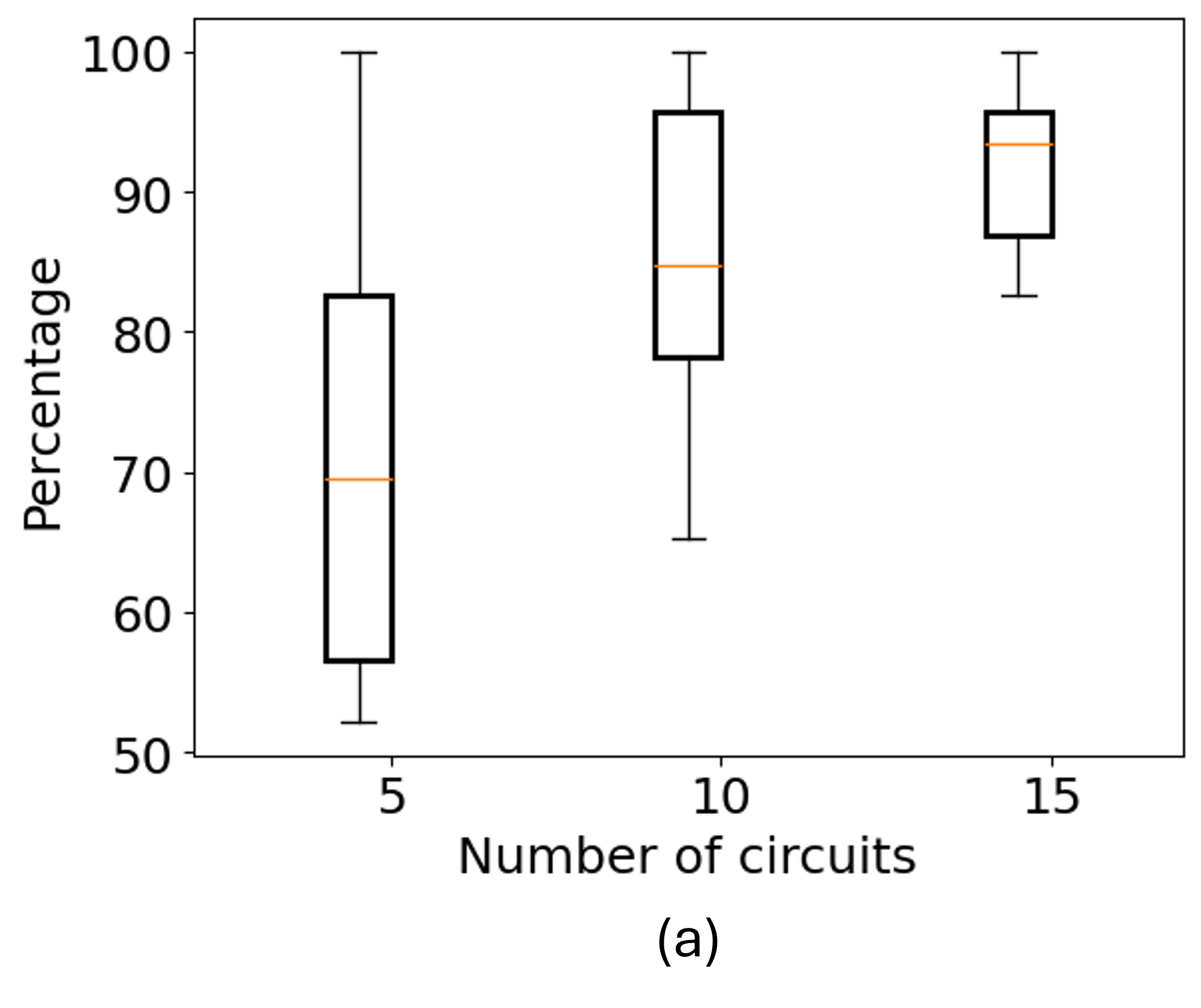}
                 \vspace{0mm}
         \label{fig:re_s_auto}
                \centering
                    \includegraphics[width=0.46\columnwidth]{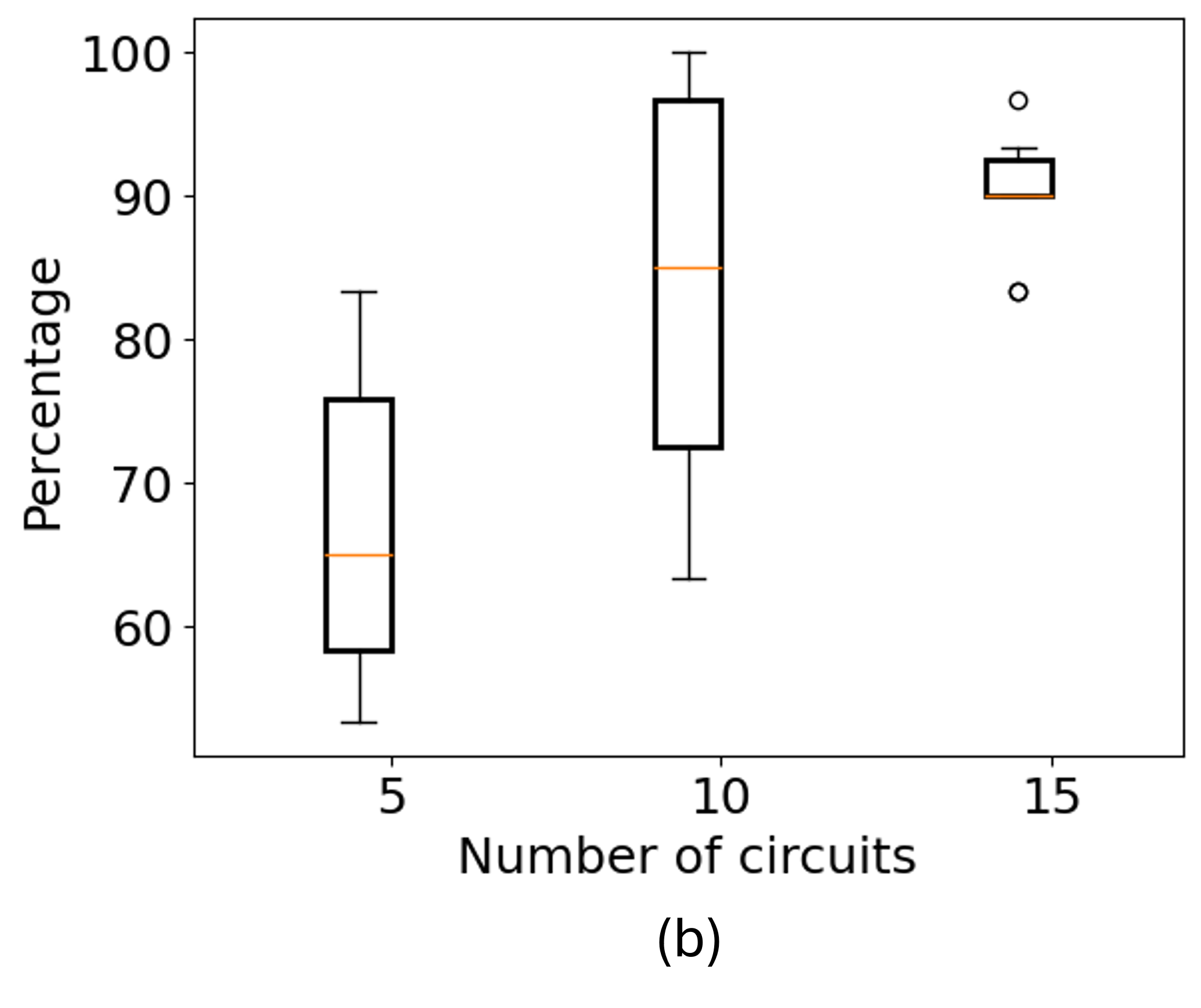}
                 \vspace{0mm}
        \label{fig:re_c_auto}
        \vspace{-1mm}
        \caption{Percentage of (a) fake\_cambridge (b) fake\_singapore backends extracted using pools of differently sized transpiled circuits. Auto-assignment of logical-physical qubit mapping by the transpiler is used for these results. }    
        \vspace{-2mm}
        \label{ckt_pool_auto}
\end{figure}

\begin{figure} 
    \centering
        \vspace{-3mm}
        \includegraphics[width=0.9\linewidth]{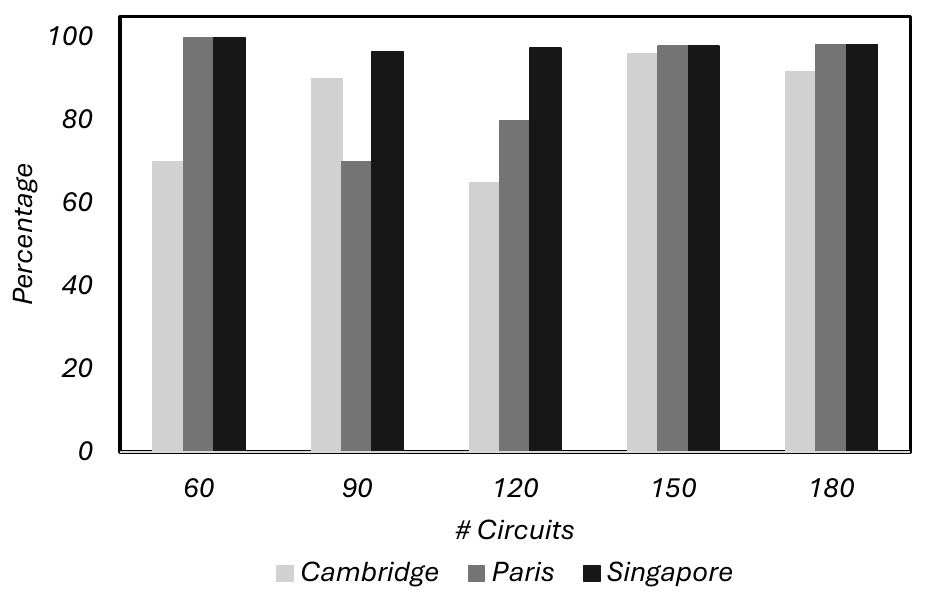}
        \vspace{0mm}
        \caption{Percentage of circuit pool that is traced accurately to its backend, for pools of various sizes.}
        \label{fig:identify}
\end{figure}

\subsubsection{Results for auto-assignment of qubits}
Next, we move on to the scenarios where the user allows the transpiler to choose the physical qubit mapping (in order to get the best possible performance vis-a-vis error rates and circuit overhead), or the user may prefer less wait time by letting the transpiler select the physical qubits among the available ones. In practice, the same backend would be in use for several circuits submitted by different users (in multi-programming environment), and thus not all physical qubits of the backend would be available for transpilation.

By default, the transpiler always chooses the coupling map with best computation fidelity and least number of SWAP gates. However, current transpilers such as Qiskit do not support multi-programming environment. To recreate the scenario where the transpiler gets a pool of programs to map to the hardware in parallel, we append the user program to another larger program (which mimics other parallel programs mapped to the same backend) and transpile on a given backend. From the logical-to-physical qubit mapping, we find out the physical qubits that are allocated for the user circuit. Then using this information, we derive the physical topology for the user circuit from the topology derived for the whole circuit. 

We transpiled several multi-qubit circuits of the same size (5 qubits, 10 qubits and 15 qubits) on the backends with auto-assignment by appending different dummy programs to the user program. From Fig. \ref{re_mixed_auto}(a) it is observed that there is a steady improvement in the percentage of the coupling map of the backend fake\_singapore predicted with an increase in the number of circuits considered, and also with the increase in the size of the circuits; as seen earlier. To derive the entire coupling map of the backend (100\%) we need as little as three 15-qubit circuits. For smaller circuits, more programs are needed. For example, eighteen 5-qubit circuits and twelve 10-qubit circuits are required to extract the backend. 
Carrying out similar experiments for fake\_cambridge backend (Fig. \ref{re_mixed_auto}(b)) we note that entire backend can be extracted using 9 15-qubit circuits, eighteen 5-qubit circuits or twelve 10-qubit circuits.
Further analysis is conducted by taking 10 randomly selected pools of 5, 10 and 15 qubit transpiled circuits, and investigating the amount of coupling map predicted correctly (Fig. \ref{ckt_pool_auto}(a) and (b)). As expected, the percentage of the map predicted increases as the size of the pool increases. This scenario took about 0.0029s for fake\_Cambridge and 0.0011s for fake\_Singapore (Table \ref{tab:table2}).

\subsubsection{Results for mixture of backends}
To make the situation even more realistic, we assume a mixture of backends and a pool of programs that are transpiled randomly on one of these backends (in conjunction with other circuits from other users that are also being executed on the same backends, as explained above). Next, we extract the backend from the pool of transpiled circuits. For this experiment, we run a set of circuits on 3 different backends - fake\_parisv2, fake\_cambridge and fake\_singapore. To trace the backend, we derive the physical qubit topology from each transpiled circuit (like in Fig. \ref{heuristic}) and check if the topology is a subgraph of any of the three backends, taking into account the exact physical qubit numbers in the derived topology and the physical coupling map of the backends. If so, we tag the program and the subgraph for that backend. We perform this experiment for pools of 60 to 180 circuits (Fig. \ref{fig:identify}). We could correctly trace back the hardware for up to 97.33\% of the transpiled circuits. For the remaining circuits, the derived physical topology is a subgraph common to more than one hardware suite (e.g. Fig. \ref{fig:cmap}). The percentage of correctly traced circuits are varies with the size of the circuit pools since each pool has different sets of transpiled circuits. For example, the pool of 60 and 90 circuits are completely different. Nevertheless, its notable that large fraction of circuits are successfully traced back to the correct backend.
The maximum percentage of correct backend prediction is 100\% for IBM\_Paris and IBM\_Singapore and 96\% for IBM\_Cambridge. 
This scenario took 0.011s for the smallest pool (60 circuits) to 0.062s for the largest (180 circuits) pool (Table \ref{tab:table2}).

\begin{table}[]
\centering
\caption{Time needed for execution}
\begin{tabular}{cc|cc}
\multicolumn{2}{c}{Backend Identification} & \multicolumn{2}{c}{Topology Extraction} \\\hline
Circuit Pool         & Time (in s)         & Coupling Map             & Time(in s)     \\\hline\hline
60                   & 0.011               & Line     & 0.24         \\
90                   & 0.024               & H-shaped     & 0.26         \\
120                  & 0.027               & Loop           & 0.21         \\
150                  & 0.049               & T-shaped           & 0.22         \\
180                  & 0.062               &                     &               \\ \hline\hline
\end{tabular}
\label{tab:table}
\end{table}

\begin{table}[]
\centering
\caption{Time needed for assembling the backend map}
\begin{tabular}{c||cc}
Backend   & Time(for auto-assignment) (in s) & Time (in s) \\\hline
Cambridge & 0.0029            & 0.0009      \\
Singapore & 0.0011            & 0.0009     \\\hline\hline
\end{tabular}
\label{tab:table2}
\end{table}

\section{Conclusion}
 Quantum backends vary in terms of noise behavior, basis gate set, coupling architecture and speed, among other parameters. The user can choose from various quantum hardware, qubit technologies, and coupling maps; however, the execution of the circuit is a black-box operation in a quantum cloud. The performance of the user program is largely dependent on the backend that was used. Moreover, the hardware scheduler may be buggy and/or third-party provider may not be trustworthy. Therefore, the quantum circuits may be executed on less efficient and/or unreliable hardware. Thus, gaining transparency using forensics on the backend will be extremely valuable. Effective forensics can facilitate many applications including establishing trust in the quantum cloud services. We introduced the problem of forensics in the domain of quantum computing where the objective is to trace the hardware and its coupling map on which the quantum program has been transpiled. We perform experiments on various coupling topologies (linear, T-shaped, H-shaped, and loop) on various IBM backends. Results indicate our ability to trace the coupling map and the backend from the transpiled circuits with high degree of accuracy. 


\section{Acknowledgment}

This work is supported in parts by NSF (CNS-1722557, CNS-2129675, CCF-2210963, CCF-1718474, OIA-2040667, DGE-1723687, DGE-1821766, and DGE-2113839) and Intel's gifts.

\vspace{12pt}


\begin{thebibliography}{00}
\bibitem{b0} https://quantum-computing.ibm.com/services/resources/docs/resources
/manage/systems/queue
\bibitem{b01}Upadhyay, Suryansh, and Swaroop Ghosh. "SHARE: Secure Hardware Allocation and Resource Efficiency in Quantum Systems." arXiv preprint arXiv:2405.00863 (2024).
\bibitem{b1} Wu, Jindi, Tianjie Hu, and Qun Li. "Detecting Fraudulent Services on Quantum Cloud Platforms via Dynamic Fingerprinting." arXiv preprint arXiv:2408.11203 (2024).
\bibitem{b2} Wu, Jindi, Tianjie Hu, and Qun Li. "Q-ID: Lightweight Quantum Network Server Identification through Fingerprinting." IEEE Network (2024).
\bibitem{b3} Smith, Kaitlin N., Joshua Viszlai, Lennart Maximilian Seifert, Jonathan M. Baker, Jakub Szefer, and Frederic T. Chong. "Fast fingerprinting of cloud-based nisq quantum computers." In 2023 IEEE International Symposium on Hardware Oriented Security and Trust (HOST), pp. 1-12. IEEE, 2023.
\bibitem{b4}K. Phalak, A. A. -. Saki, M. Alam, R. O. Topaloglu and S. Ghosh, "Quantum PUF for Security and Trust in Quantum Computing," in IEEE Journal on Emerging and Selected Topics in Circuits and Systems, vol. 11, no. 2, pp. 333-342, June 2021, doi: 10.1109/JETCAS.2021.3077024.

\bibitem{b5} Robert Wille, Daniel Große, Lisa Teuber, Gerhard W. Dueck, Rolf Drechsler: RevLib: An Online Resource for Reversible Functions and Reversible Circuits. Int'l Symp. on Multi-Valued Logic, 2008
\bibitem{b12} Upadhyay, S., \& Ghosh, S. (2024). SHARE: Secure Hardware Allocation and Resource Efficiency in Quantum Systems. https://arxiv.org/abs/2405.00863
\bibitem{b13}Karie, N. M., \& Venter, H. S. (2015). Taxonomy of challenges for digital forensics. Journal of Forensic Sciences, 60(4), 885–893.
\bibitem{b14} Sayakkara, A., Le-Khac, N.-A., \& Scanlon, M. (2019). A survey of electromagnetic side-channel attacks and discussion on their case-progressing potential for digital forensics. Digital Investigation, 29, 43–54.
\bibitem{b15} Rekhis, S., \& Boudriga, N. (2011). Logic-based approach for digital forensic investigation in communication networks. Computers \& Security, 30(6–7), 376–396.

\bibitem{b16} Thongkamwitoon, T. (2014). Digital forensic techniques for the reverse engineering of image acquisition chains. Imperial College London.
\end{thebibliography}
\end{document}